\documentclass[10pt,a4paper]{article}
\usepackage{amsmath}
\usepackage{amsfonts}
\usepackage{amssymb}
\usepackage{graphicx}
\usepackage{epsfig}
\begin{document}
\title{A thermodynamic switch for chromosome colocalization}
\author{Mario Nicodemi$^{a,b}$(*), Barbara Panning$^c$,\\ and Antonella Prisco$^d$\\
$^a$ Department of Physics and Complexity Science, University\\ of Warwick, UK, 
and $^b$ INFN Napoli, Italy\\
$^c$ Department of Biochemistry and Biophysics,\\ University of California
San Francisco, California, USA\\
$^d$ CNR Inst. Genet. and Biophys. `Buzzati Traverso',\\ 
Via P. Castellino 111, Napoli, Italy}
\date{\today}
\maketitle

A general model for the early recognition and colocalization of homologous DNA sequences is proposed. We show, on a thermodynamic ground, how the distance between two homologous DNA sequences is spontaneously regulated by the concentration and affinity of diffusible mediators binding them, which act as a switch 
between two phases corresponding to independence or colocalization of pairing regions. 

\bigskip

Chromosome recognition and pairing is a general feature of nuclear
organization. 
In particular, these phenomena have a prominent role (and are comparatively better studied) 
in meiosis, the specialized cell division necessary for the production
of haploid gametes from diploid nuclei. During the prophase of the first meiotic division, homologous chromosomes identify each other and pair via a still mysterious long-distance reciprocal recognition process 
\cite{Kleckner98,ScottHawley05,Zickler06}.

Many hypotheses exist on the mechanisms underlying the early stages of coalignment of homologs along their length (see ref.s in \cite{Kleckner98,ScottHawley05,Zickler06}). 
A longstanding idea is that pairing may occur via unstable interactions, such as a direct physical contact between DNA duplexes (the ``kissing model", see, e.g., \cite{KlecknerWeiner93}). 
Pairing initially based on non permanent interactions has the important 
advantage of preventing ectopic association between non-homologous chromosomes, and avoid topologically unacceptable entanglements, leaving space to adjustments
\cite{KlecknerWeiner93}. 
Several mechanisms could contribute to the outcome of the process, 
e.g., costrained motion of chromosome in territories, bouquet formation at telomeres, tethering to the nuclear envelope. While chromosome full alignment includes several stages, the early physical contact and colocalization could be driven by specific chromosomal regions bridged by molecular mediators.
In this complex scenario, though, the crucial question on the mechanical origin of  early recognition and pairing remains unexplained.

Here we explore the thermodynamic properties of a recognition/pairing mechanism based on weak, biochemically unstable interactions between specific  DNA sequences and molecular mediators binding them. We show that randomly diffusing molecules can produce a long-distance interaction mechanism whereby homologous sequences 
spontaneously recognize and become tethered to each other. This colocalization mechanism is tunable by two ``thermodynamic switches", namely the concentration of molecular mediator and their affinity for their binding sites. When threshold 
values in the concentration, or affinity, of mediators are exceeded, homologous sequences are joined together, else they move independently.

\bigskip

\begin{figure}
\hspace{-4.5cm}
\epsfig{figure=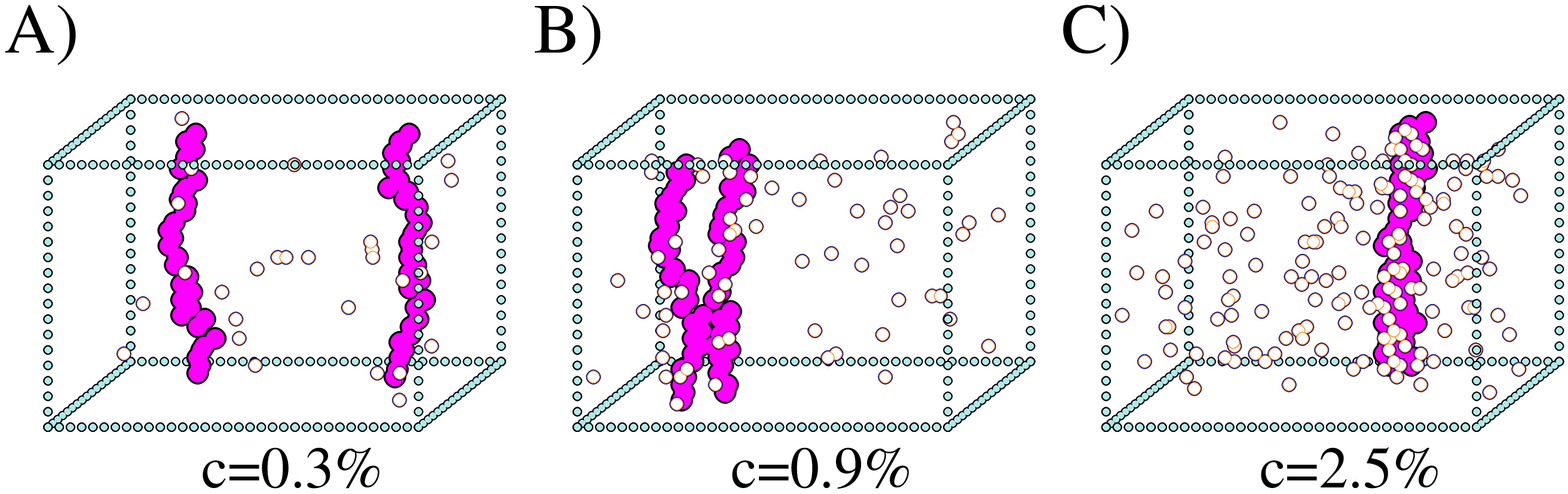,width=10cm,angle=-90}
\vspace{1cm}
\caption{\label{transitionf} 
Pictures of typical configurations, from computer simulations, of the model system at thermodynamic equilibrium, 
in the two described phases discussed in Fig.\ref{transition} (panel {\bf A},  independent motion; panel {\bf C}, colocalization) and  their  intermediate crossover region (panel {\bf B}), for the shown values of the concentration of molecular mediators, $c$ (here $E_X=1.2kT$).
}
\end{figure}

{\bf Model: } 
Our model includes (see Fig.\ref{transitionf}) two homologue segments involved in mutual recognition and pairing,  described as a self-avoiding bead chains, a well established model of polymer physics \cite{EdwardsDoi}, and a concentration, $c$, of Brownian molecular factors having a chemical affinity, $E_X$, for them. 
We investigate the thermodynamics properties of the system by Monte Carlo (MC) computer simulations \cite{Binder}. 
For computational purposes, chromosomal segments and molecules are placed in a volume consisting of a cubic lattice with spacing $d_0$ (our space unit, of the order of the molecular factors length) and linear sizes $L_x=2L$, $L_y=L$ and $L_z=L$ (see Fig.\ref{transitionf}). In each simulation, the `beads' of the chromosomal segments start from a straight, vertical line configuration, at a distance L from each other, and molecular mediators from a random initial distribution. Diffusing molecules randomly move from one to a nearest neighbor vertex on the lattice. On each vertex no more than one particle can be present at a given time. The chromosomal segments diffuse as well on such a lattice performing a Brownian motion under the constraint that two proximal `beads' on the string must be  within a distance $\sqrt{3}d_0$ from each other (i.e., on next or nearest next neighboring sites on the lattice). For the sake of simplicity, we disregard here the rest of the chromosomes and 
DNA segment ends are costrained to move tethered to the bottom and top plane of the system volume (Fig.\ref{transitionf}). 
When neighboring a chromosomal chain, molecules interact with it via a binding energy $E_X$. Below, we mainly discuss the case where $E_X$ is of the order of a ``weak'' hydrogen bond-like energy, say 3 kJ/mole, which
at room temperature corresponds to $E_X=1.2kT$ \cite{Watson}. In our simulations, at each time unit (corresponding to a MC lattice sweep) the probability of a particle to move to a neighboring empty site is proportional to the Arrhenius factor $r_0\exp(-\Delta E/kT)$, where $\Delta E$ is the energy barrier in the move, $k$ the Boltzmann constant and $T$ the temperature \cite{Watson,Stanley}. The factor $r_0$ is the reaction kinetic rate, depending on the nature of the molecular factors and of the surrounding viscous fluid, and sets the time scale. We employ $r_0=30 sec^{-1}$, a typical value in biochemical kinetics. Averages are over up to 2048 runs from different initial configurations. 

\bigskip

{\bf Results: } 
First we show how the interaction of chromosomes with molecular mediators drives  colocalization. 
To this aim, we calculated the thermodynamic equilibrium value of the average square distance (relative to the system linear size $L$) between the two chromosomal segments: 
\begin{equation}
d^2=\frac{1}{N}\sum_{z=1}^N {\langle r^2(z)\rangle \over L^2}
\end{equation}
where $N$ is the number of beads in each string (here $N=L$) and 
$\langle r^2(z)\rangle$ is the average (over MC simulations) 
of the square distance of the beads at `height' $z$.
The average value of $d^2$ is maximal when the two `chromosomes' float independently and decreases if  parts of the polymers become colocalized, approaching zero when a perfect alignment is attained. 

\begin{figure}
\hspace{0cm}
\epsfig{figure=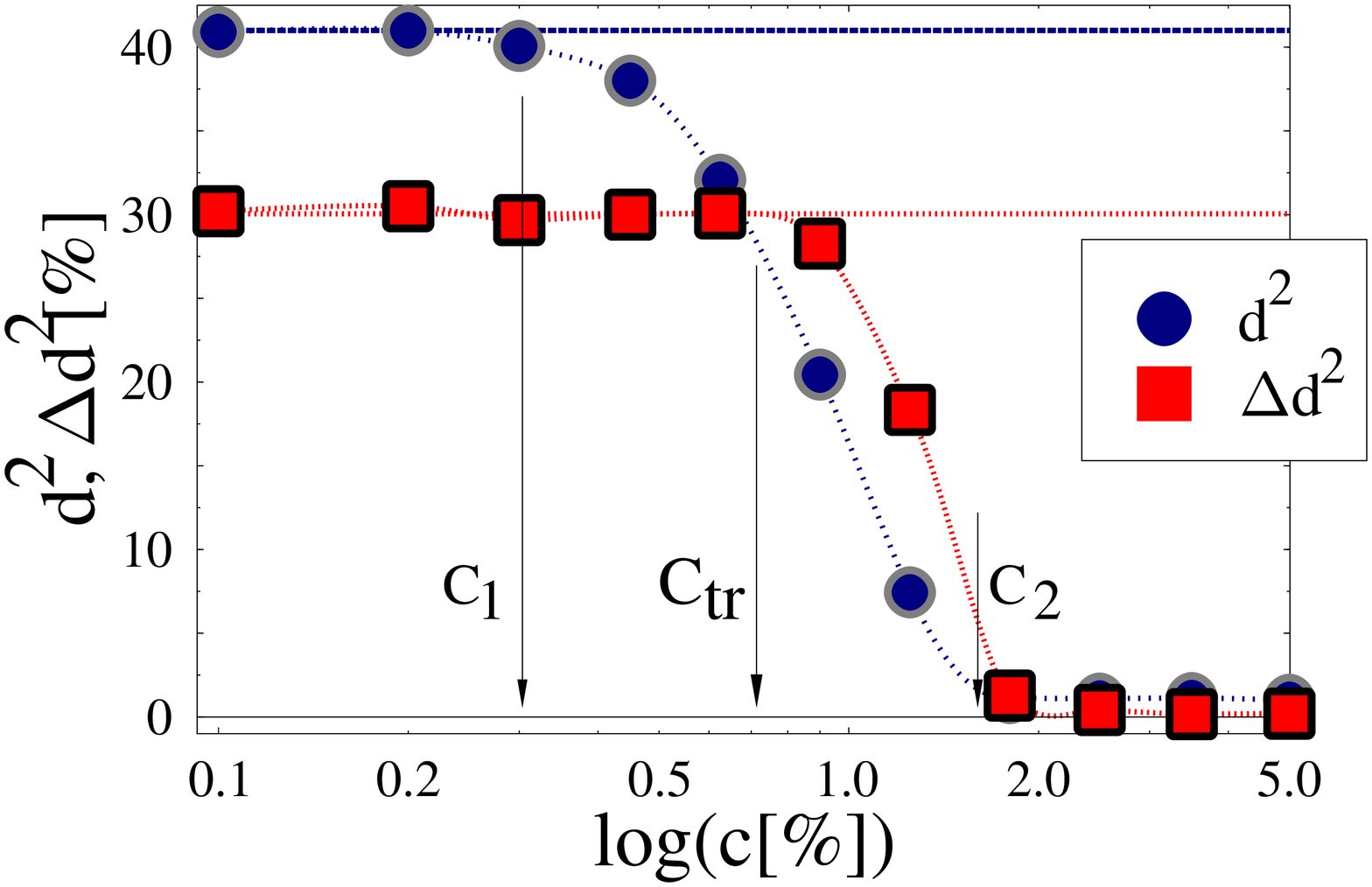,width=10.5cm,angle=-90}

\vspace{-3cm}
\hspace{0cm}
\epsfig{figure=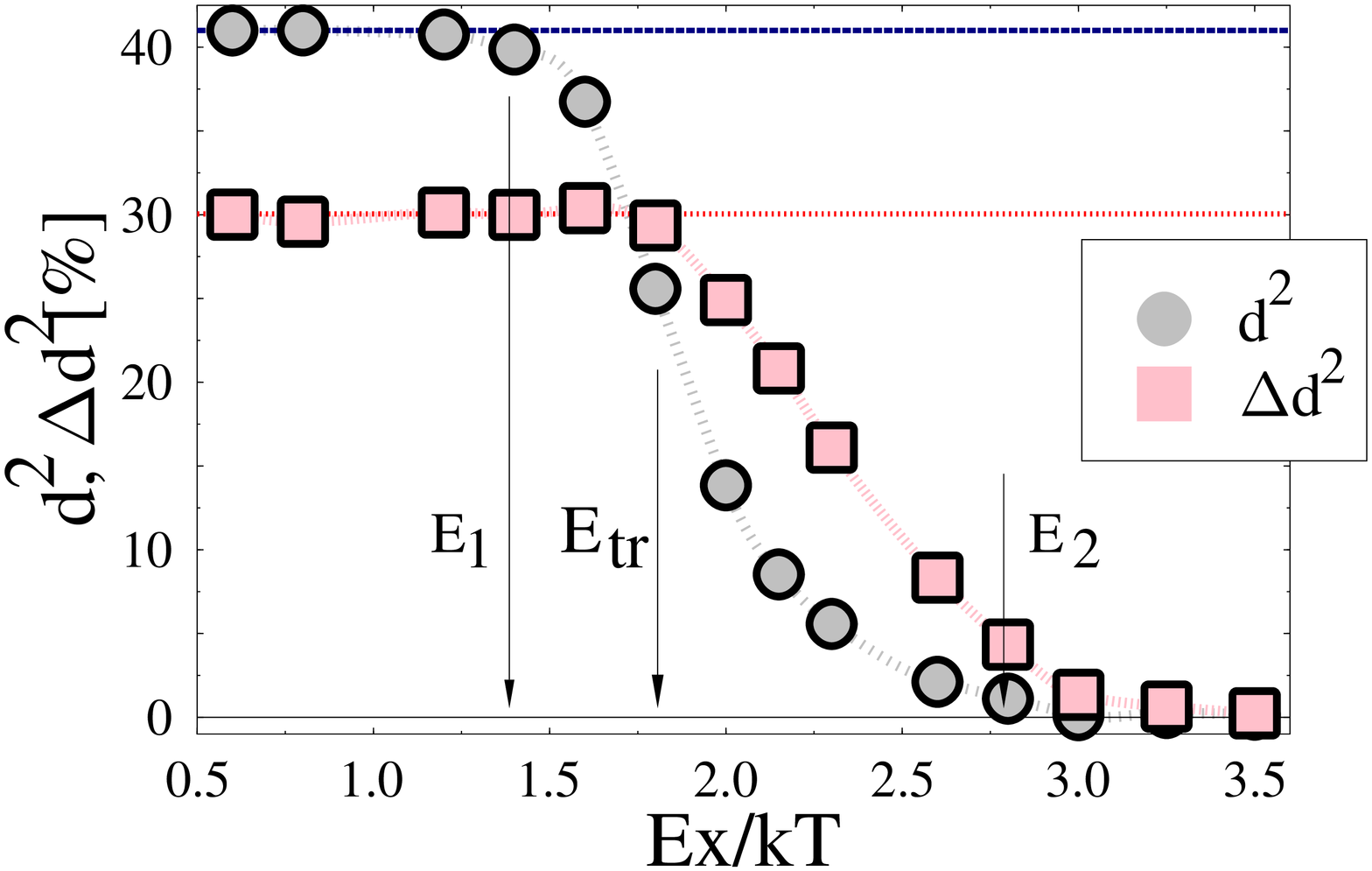,width=10.5cm,angle=-90}
\vspace{-2.5cm}
\caption{\label{transition} 
{\bf Top panel } The equilibrium chromosome average square distance, $d^2$, is shown as a 
function of the concentration of binding molecules, $c$ (here the 
molecule/chrom. affinity is $E_X=1.2kT$):
for $c<c_{tr}\simeq 0.7\%$, $d^2$ approaches values as big as the 
system size and chromosomes are randomly and independently 
diffusing (horizontal dotted lines give the values found for pure 
random walks); 
for $c>c_{tr}$, $d^2$ rapidly decays to zero, showing that they have
colocalized. Around $c_{tr}$ there is a crossover regime, 
approx. between $c_1$ 
and $c_2$, 
where chromosomes tend to align since $d^2$ is smaller than in the region
where they move independently, but its fluctuations, $\Delta d^2$, 
are of the order of $d^2$; here chromosomes are only transiently colocalizing. 
{\bf Bottom panel } A similar behaviour is found when $d^2$ is plotted as a function of the chemical affinity, $E_X$, shown here for $c=0.1\%$. 
}
\end{figure}

The equilibrium distance, $d^2$, depends on the concentration, $c$, of mediators. 
At low concentration (see Fig.\ref{transition}, e.g., $c<c_1$) $d^2$ has a value of the order of the system size (around $40\%$ of $L^2$), corresponding to the expected average distance of two independent strings undergoing Brownian motion in a box of size $L$; a typical configuration for $c=0.3\%$ being shown in Fig.\ref{transitionf} panel A). 
Indeed, the physical basis for the independence of chromosomes exposed to a  low concentration of mediating molecules is intuitive:  pairing can 
occur when
bridges are formed by molecules attached to couples of binding sites. A single bridging event, however, can be statistically quite unlikely since `weak' bonds are biochemically unstable and to form a bridge a diffusing molecule must first find (and bind) a site on one chromosome and then together they have to successfully encounter the second one. 

Fig.\ref{transition} shows, however, that when $c$ is higher than a threshold value, $c_{tr}$ (for $E_X=1.2kT$, $c_{tr}\simeq 0.7\%$), $d^2$ collapses to zero: this is the sign that the two `chromosomes' have colocalized; 
a typical picture of the system state, for $c=2.5\%$, is shown in panel C) of Fig.\ref{transitionf}. 
Actually, when $c$ is high enough chances increase to form multiple bridges and, as they reinforce each other, configurations where molecules hold together the two polymers become stabilized. The threshold concentration value, $c_{tr}$, corresponds to the point where such a positive mechanisms becomes winning, and can be approximately defined by the inflection point of the curve $d^2(c)$. 
Alike phase transitions in finite-size systems \cite{Binder,Stanley} (see below), around $c_{tr}$ there is a crossover region which can be located, for instance, between the concentrations $c_1$ and $c_2$ (see Fig.\ref{transition}) 
defined by the criterion that $d^2$ is close within $5\%$ to the random or zero plateau value 
(for $E_X=1.2kT$, $c_1\simeq 0.3\%$ and $c_2\simeq 2\%$). 

In Fig.\ref{transition}, along with the distance between chromosomes, $d^2$, we 
plot the squared fluctuations of the distance (i.e., its statistical variance), 
$\Delta d^2(c)$, as a function of the concentration of mediators. 
For $c<c_1$, both $d^2(c)$ and $\Delta d^2(c)$ have the
non zero value found for non interacting Brownian strings in the independent 
diffusion regime ($\Delta d^2\sim 30\%$); instead, $\Delta d^2(c)=0$ for 
$c>c_2$ in the tight colocalization regime. Interestingly, in the 
crossover region, $d^2(c)$ is smaller than in the purely random regime, 
although it has marked fluctuations ($\Delta d^2(c)$ can be even 
larger than $d^2(c)$). This situation is illustrated by a picture
of a typical configuration, for $c=0.9\%$, shown in panel B) 
of Fig.\ref{transitionf}. In such an intermediate regime
chromosome couples are continuously formed and disrupted.

Summarizing, our results show that colocalization is spontaneously induced by the `collective' binding of molecular mediators and occurs only when $c$ is above a critical value, $c_{tr}$, i.e., in the `colocalization phase'. Conversely, when $c$ is  below $c_{tr}$, $d^2(c)$ has the same value found for two non interacting Brownian strings. This is the `random phase', where chromosomes are independent. The concentration of mediators acts as a switch between the two phases, while around the critical threshold chromosomes undergo transient interactions. 

A similar effect is found when, for a given (high enough) concentration, $c$, the chemical affinity, $E_X$, of binding sites is changed (see Fig.\ref{transition} lower panel): when $E_X$ is smaller than a threshold value, $E_{tr}$, the two polymers float independently one from the other. Around $E_{tr}$ a crossover region is found, and as soon as $E_X$ gets larger than $E_{tr}$, an effective attraction between polymers is established and they are  spontaneously colocalized. 
Another potential layer of regulation of the system is the number of binding sites for molecular mediators. In fact, a reduction in the number of binding sites produces the same effect of a reduction in the affinity of mediators, that is,  chromosomes become unable to find and bind each other.

\begin{figure}
\centerline{\hspace{-6cm}
\epsfig{figure=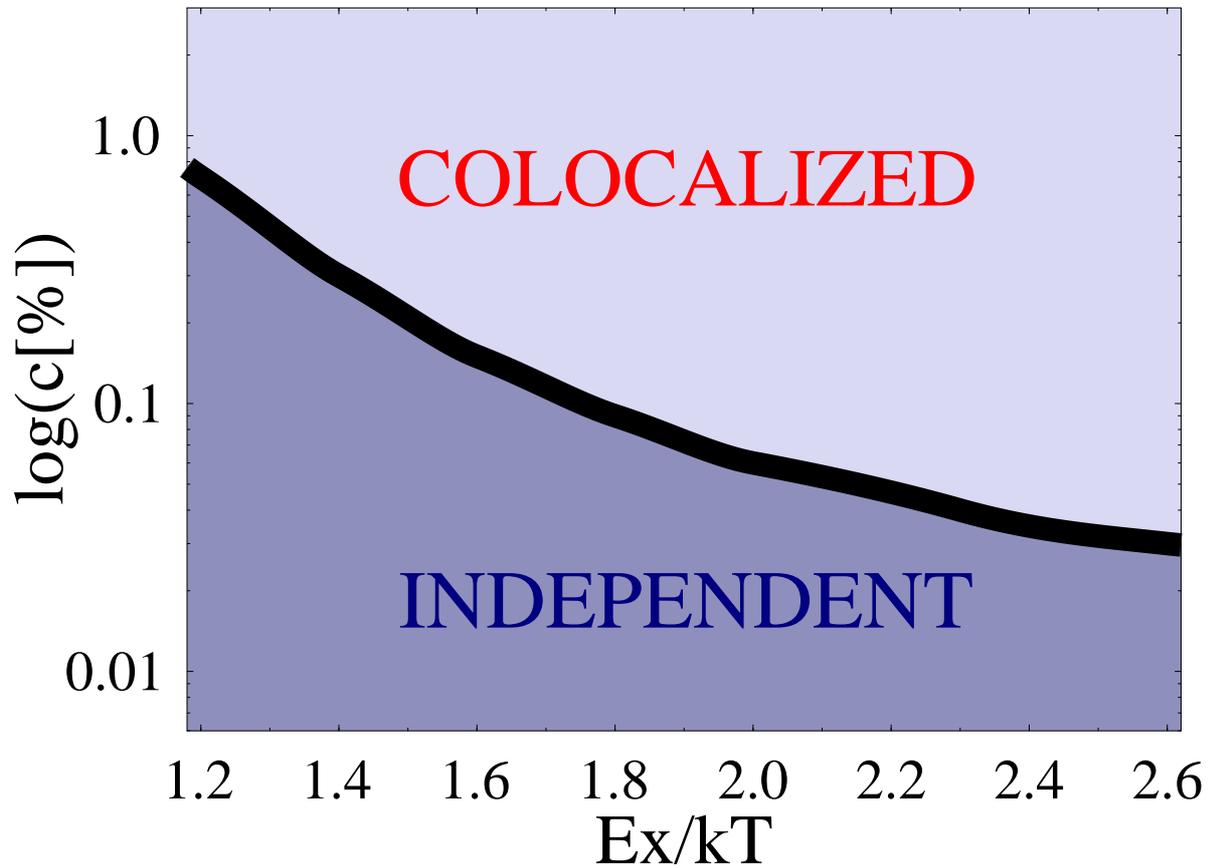,width=15cm,angle=-90}}
\caption{\label{ph_diag} 
This phase diagram shows the state of the two chromosomes at 
thermodynamic equilibrium in a range of values of chemical
affinity and concentration of their molecular mediators, i.e., 
in the $(E_X,c)$ plane. For small $E_X$ and $c$, chromosomes move
independently while, above a transition region, they spontaneously
colocalize. The transition line, $c_{tr}(E_X)$, is marked by the heavy
black line. Colocalization, thus, can be spontaneously attained by 
upregulation of mediator concentration, $c$, or of molecule chemical affinity, $E_X$, to chromosomal sequences. 
}
\end{figure}

The pairing mechanisms illustrated above has a thermodynamics origin. It is a `phase transition' \cite{Stanley}  occurring when entropy loss due to polymer colocalization is compensated by particle energy gain as they bind both polymers, the lower $E_X$ the higher the concentration, $c$, required. Actually, the transition is found in a broad region of the $(E_X,c)$ plane, as shown in Fig.\ref{ph_diag} where the system phase diagram is plotted
in a range of typical biochemical values of ``weak'' binding energies $E_X$. For very low values of $E_X$ the colocalization can be, instead, impossible. The overall properties of such a phase diagram (independent v.s. colocalized chromosomes) are robust to changes in the model details, though the precise location of the different phases can be affected \cite{Stanley}. 
Summarizing, when soluble mediators bind a specific recognition sequence on homologous chromosomes, recognition and colocalization of homologs can occur, as a result of a robust and general thermodynamic phenomenon, namely a phase 
transition occurring in the system.
The higher the affinity of mediators for chromosomal binding sites, the lower is the threshold concentration of mediators that promotes colocalization (see Fig.\ref{ph_diag}).

\bigskip

{\bf Discussion: } 
We described a general colocalization mechanism, grounded on thermodynamics, whereby specific regions of a pair of chromosomes can spontaneously recognize each other and align. Physical juxtaposition is mediated by sequence-specific molecular factors that bind DNA via weak, non permanent, biochemical interactions.  When the concentration/affinity of molecular mediators is above a critical threshold 
an effective attraction between their binding regions is generated, leading to a close alignment; else chromosomes float away from each other by Brownian motion. In the threshold crossover region, pairing sites undergo transient interactions: the average distance is shorter than in the purely random regime, but marked fluctuations are observed. 

In our simulations, the two homologous pairing regions are described as polymers diffusing with their ends tethered to the upper and lower planes of the system box. This recalls telomeres tethering to the nuclear envelope observed at meiosis. While it is not a prerequisite for the switch mechanism, on the other hand, it can enhance the switch effects \cite{Kleckner98,ScottHawley05,Zickler06}. 
Releasing such a constraint doesn't change the general results, but pairing regions would collapse in a more disordered geometry. The overall properties of the phase diagram (independent vs. colocalized chromosomes) are robust to changes in the model details \cite{noinpreparation}. A model including many a pair of chromosomes has longer equilibration times, as expected in a crowded environment, yet, its phase diagram is unchanged. The scenario is also unaltered in the case of mediators that interact with each other and aggregate. 

An implication of this model is that a cell can regulate the initiation of homologous chromosome interaction by up-regulating the concentration of mediators or their affinity for DNA sites (e.g., through changes in the chromatin or by a chemical modification of the mediator).
This switch has general and robust roots in a thermodynamics phase transition \cite{Stanley}, irrespective of ultimate molecular and biochemical basis. 
In real cells, specific short chromosomal regions (``pairing centers") could mediate the early steps of homolog recognition, and act as a seed and reference point to a subsequent stable long scale chromosomal pairing, which could involve additional mechanisms.
A speculation is that the threshold effect can be exploited to ensure a precise control of pairing formation/release, while the presence of a crossover region in concentration to reduce undesired entanglements.
The initial binding molecules could, in turn, help the sequences in recruiting complexes later used to other purposes (e.g., in pairing stabilization, synapsis, recombination). 

In the present model individual mediators do not need to be strongly binding to glue homologous chromosomes together, and any molecules with above threshold affinity can induce attraction. Specificity of colocalization among many chromosome pairs could be, indeed, obtained by sets of molecules binding, with higher affinities, specific homologous sequences. 
While the molecular mediators considered here are supposed to have more than one ``DNA binding domain", proteins that can bind a single DNA site, but are able to make protein-protein interactions, could also mediate co-localization. 
As a pair of linked proteins is, in fact, a single molecular mediator the thermodynamics picture is unchanged. Finally, direct DNA duplex interactions \cite{KlecknerWeiner93} could replace, or help, binding molecules. A duplex kissing site would correspond in our model to a binding site with a molecular mediator already attached, so the overall behavior should be similar. 

Experimental discoveries on meiotic pairing have accomplished huge progresses, but the mechanisms for homologue early coalignment are still unclear \cite{Kleckner98,ScottHawley05,Zickler06}. 
In {\em C. elegans}, for instance, homologs proper pairing is primarily regulated by special telomeric regions, known as ``pairing centers" (PCs) \cite{McKim88,Villeneuve94,MacQueen05}. 
Homologous PCs interact, during early prophase, with HIM/ZIM Zn-finger proteins which are necessary to mediate pairing \cite{Phillips05,Phillips06}. Specific sites and proteins are also involved in meiotic pairing of {\em Drosophila}. In male, on the X and Y chromosomes, a 240bp repeated sequence in the intergenic spacer of  rDNA acts as a pairing center, and autosomes 
pair, as well, by the interaction of a number of 
sites (see ref.s in \cite{ScottHawley05,Zickler06}). A similar behavior is observed in {\em Drosophila} female 
\cite{Hawley92,Dernburg96}. In {\em Drosophila} males, special proteins, SNM and MNM, have been also discovered which bind X-Y and autosomal pairing sites at prophase I, and are required for pairing \cite{Thomas05}. 
The question is open whether the present model applies to such an experimental scenario. 
In a picture where pairing is mediated by unstable interactions, thermodynamics dictates, anyway, a precise framework showing that minimal ``ingredients", such as soluble DNA binding molecules and homologous arrays of binding sites, can in fact be sufficient for pairing if the balance of mediator concentration and DNA affinity is appropriate. 

Our thermodynamic switch theory is prone to experimental tests (e.g., the existence of threshold effects in mediator concentration, $c$). It can be exploited, as well, for a quantitative understanding of the effects on pairing, e.g., of deletions (which can be modeled here by reducing the binding site number within $L$), or of chemical modifications of binding sequences (modeled by changes in $E_X$), and to guide the search for candidates for chromosomal sites and interaction mediators. 
Finally, the general message of the model may be applicable to various cellular processes that involve the spatial reorganization of DNA in nuclear space (e.g.,  organization of chromosomal loci and territories, justapposition of DNA sequences in transcriptional regulation, somatic pairing, pairing of X chromosomes at the onset of X inactivation
\cite{Kleckner98,ScottHawley05,Zickler06,Lee06,ap14,mario,nicodemi07b,Lanctot,Misteli07}).

\bigskip

We thank N. Kleckner and A. Storlazzi for very helpful discussions and critical reading of the manuscript.

\newpage

\end{document}